\documentclass{article}

\usepackage{graphicx}
\usepackage{amssymb, amsmath,amsthm}
\usepackage{enumerate}
\usepackage{mathrsfs}

\usepackage{cite}      % IEEE rec

\newcommand{\dual}{\hat}
\newcommand{\wdual}{\widehat}
\newcommand{\trans}{\check}

\newcommand{\gv}{\mathbf{g}}
\newcommand{\hv}{\mathbf{h}}
\newcommand{\pv}{\mathbf{p}}
\newcommand{\qv}{\mathbf{q}}

\newcommand{\rhov}{\mathbf{\rho}}
\newcommand{\ps}[2]{\psi^{(#1)}(#2)}

\newcommand{\f}[2]{f^{(#1)}(#2)}
\newcommand{\ft}[2]{\trans{f}^{(#1)}(#2)}

\newcommand{\path}{\mathsf{p}}
\newcommand{\ZZ}{\mathbb{Z}}
\newcommand{\RR}{\mathbb{R}}
\newcommand{\CC}{\mathbb{C}}
\newcommand{\upstar}{{}^{*} \!}

\newcommand{\pep}{Y_{G,m}}
\newcommand{\diag}{\operatorname{diag}}
\newcommand{\leaves}{\mathscr{L}}
\newcommand{\tree}{\mathscr{T}}
\newcommand{\arxiv}[1]{#1}
\newcommand{\noarxiv}[1]{}

\newtheorem{thm}{Theorem}[section]
\newtheorem{obs}[thm]{Observation}
\newtheorem{lem}[thm]{Lemma}

\newtheorem{prop}[thm]{Proposition}
\newtheorem{defn}[thm]{Definition}

\title{Fourier transform inequalities for phylogenetic trees}
\author{Frederick A. Matsen\\
 Department of Statistics\\
 University of California, Berkeley\\
 367 Evans Hall \#429\\
 Berkeley, CA 94720-3860\\
 USA\\
 http://www.stat.berkeley.edu/$\tilde{\ }$matsen/ }

\begin{document}
\maketitle

\begin{abstract}
Phylogenetic invariants are not the only constraints on site-pattern
frequency vectors for phylogenetic trees. A mutation matrix, by its
definition, is the exponential of a matrix with non-negative
off-diagonal entries; this positivity requirement implies non-trivial
constraints on the site-pattern frequency vectors. We call these
additional constraints ``edge-parameter inequalities.'' In this paper,
we first motivate the edge-parameter inequalities by considering a
pathological site-pattern frequency vector corresponding to a quartet
tree with a negative internal edge.  This site-pattern frequency
vector nevertheless satisfies all of the constraints described up to
now in the literature. We next describe two complete sets of
edge-parameter inequalities for the group-based models; these
constraints are square-free monomial inequalities in the Fourier
transformed coordinates. These inequalities, along with the
phylogenetic invariants, form a complete description of the set of
site-pattern frequency vectors corresponding to \emph{bona fide}
trees. Said in mathematical language, this paper explicitly presents
two finite lists of inequalities in Fourier coordinates of the form
``monomial $\leq 1$,'' each list characterizing the phylogenetically
relevant semialgebraic subsets of the phylogenetic varieties.
\end{abstract}

\noarxiv{
}

\section{Introduction}
\noarxiv{ \PARstart{T}{he} }
\arxiv{ The }
Bayesian and maximum-likelihood methods in phylogenetics can be
classified as ``model based.'' That is, at some stage in the analysis,
one assumes a mutation model and calculates the likelihood of the
observed data for a given tree and set of model parameters. We will
call the set of site-pattern frequency vectors generated on a fixed tree by a mutation
model under legal parameter settings a ``tree image.'' One of the main
goals of the emerging field of phylogenetic geometry
\cite{allmanRhodesPhyloInvarsGM03,casanellasFSK3P07,
cavenderFelsensteinSimpleInvars87, kimOranges00,
sturmfelsSullivantToric05} is to locate these tree images in site
pattern frequency space.  Such work is foundational to understanding
when model-based phylogenetics does and does not succeed.

The mutation models for sequences evolving on a tree are typically
given in terms of nucleotide mutation models, which are stochastic
matrices giving the probability of various mutations at an arbitrary site.
One such matrix is associated with each edge; consequently one
multiplies matrices along paths in the tree to get the mutation matrix
along that path. Because a series of matrix multiplications is
polynomial in the entries of the matrices, one can consider the tree
image as a subset of an affine variety. 

It is then natural to apply the well-developed tools of algebraic
geometry to analyze these varieties. In particular, there has been a
flourishing of interest in the corresponding ideals of these
varieties; in the present setting these are called ``phylogenetic
invariants'' \cite{allmanRhodesPhyloInvarsGM03,
allmanRhodesIdealsVarieties07, cavenderFelsensteinSimpleInvars87,
felsensteinInferringPhylogenies04, sturmfelsSullivantToric05}.
Although not completely understood for all models, a considerable
amount of beautiful work has been done on these invariants; a very
nice overview has been published in
\cite{allmanRhodesInvariantsOverview07}.

One can then formulate a constrained optimization problem 
by optimizing the likelihood function across the
set of site-pattern frequency vectors constrained to satisfy the
phylogenetic invariants. This is the view taken by
\cite{hostenEASolvingLikelihood05} (equation (3)) where it is called
the maximum likelihood problem. Another article
\cite{sturmfelsSullivantToric05} says ``exact computation
of maximum likelihood estimates\dots can be formulated\dots as a
constrained optimization problem where the probabilities are the
decision variables and the phylogenetic invariants are the
constraints.'' A similar statement has been made in a review article
concerning the use of phylogenetic invariants for tree reconstruction
\cite{erikssonUsingInvariants07}.

These statements may be confusing for computational biologists
thinking of phylogenetic trees as descriptions of mutational processes
occurring in the evolutionary
past. Indeed, there are solutions to the phylogenetic invariants sitting in
the probability simplex which do not correspond to any reasonable
assignment of branch lengths (or, more generally, edge parameters) to
a tree. In the language of algebraic geometry, the tree image is not
equal to its Zariski closure intersected
with the probability simplex.  This observation is not original to
this paper: the
authors of \cite{casanellasFSK3P07} define a useful notion of ``biologically meaningful''
solutions to the phylogenetic invariants. Their criterion is satisfied if the Fourier transform of the
mutation matrices have non-negative diagonal entries.  Positivity of
Fourier transforms is indeed a
necessary condition for a mutation matrix to come from a model (see
Observation~\ref{obs:FTpos}), but is not sufficient as we demonstrate
below in our motivating example.

Our simple observation is this: \emph{mutation matrices
are the result of a continuous time Markov process operating for some
non-negative period of time.} This fact is implicit in any description of
mutation as a process in terms of rates, for example in the original
description of the Kimura models \cite{kimuraModels80}. In the
notation of Markov processes, 
\begin{equation}
  P^{(e)} = \exp \left( t_e Q^{(e)} \right)
  \label{eq:PisExpQ}
\end{equation}
where $P^{(e)}$ is the mutation matrix for an edge $e$, $t_e \geq 0$ is
elapsed time, and $Q^{(e)}$ is the mutation rate matrix. 
In this setting $Q^{(e)}$ must be a ``$Q$-matrix'', i.e. have non-negative
off diagonal entries and zero row sums \cite{stroockIntroductionMarkov05}. 

The observation (\ref{eq:PisExpQ}) implies a collection of
nontrivial square-free monomial inequalities in the Fourier
transformed probability space which ensure that a solution to a
complete set of phylogenetic invariants indeed corresponds to a
\emph{bona fide} tree. This paper develops a complete set of such
inequalities; we call them ``edge-parameter inequalities.''

First we present a very simple motivating example on the quartet tree
to illustrate the need for edge-parameter inequalities. This example
has a negative internal branch length, or, said another way, the
mutation rate matrix along that edge contains negative off-diagonal
entries. Despite this nonsensical setup, the associated site-pattern
frequency vector satisfies the phylogenetic invariants and sits in the
probability simplex.  Furthermore, the parameters satisfy the useful
``biologically meaningful'' criterion of \cite{casanellasFSK3P07},
which as noted is necessary but not sufficient for a
tree to have positive edge parameters. For our example we assume the
two-state symmetric (CFN) model with uniform distribution at the root,
labeling the two states $0$ and $1$. In the CFN model, there is only a
single parameter per edge, called the branch length. It is the amount
of time which we allow our binary Poisson mutation process to run,
thus the probability that the endpoints of an edge are in different
states is $0.5 ( 1- \exp(-2 \gamma ))$ for an edge of length $\gamma$.
Let $\theta = \exp(-2 \gamma)$; the Fourier transform
\cite{sempleSteelPhylogenetics03} of the mutation matrix of length
$\gamma$ is thus $\diag(1,\theta)$. 

Our motivating example is as follows: consider the tree on taxa
$1,2,3$ and $4$ with the $12|34$ split.  Make each pendant edge of length
$\gamma$ and internal edge of length $-\gamma$. Thus formally, by the
above, the off-diagonal entries of the mutation rate matrix for the
internal edge will be negative.  We now show that if $\gamma >
0.60938$ then the expected site-pattern frequency vector for this tree
will satisfy all of the restrictions described up to now in the
literature.

With the above notation, the nontrivial entry of the Fourier transform
of the mutation matrix
will be $\theta$ for the pendant edges and $\theta^{-1}$ for the
internal edges. In this and the following sections, we use $\pv$ to
denote points of the probability simplex and $\qv$ to denote points of
the Fourier transform of the probability simplex. We will call the
$\pv$ ``site-pattern frequency vectors'' and the
image of the probability simplex under the Fourier transform
``$q$-space.'' We
will index $\pv$ and $\qv$ with taxon state vectors $\gv$.

We use Hadamard conjugation to compute $\qv$ for the pathological
tree. The formulation for general group-based models is given in
(\ref{eq:hadProd}), but for the CFN model the calculation of $\qv$ is
quite simple. To find a given $q_{\gv}$, first let $S_\gv$ be the set
of all taxa in state $1$ according to $\gv$. Second, let $E_\gv$ be
the set of edges
in the (unique) collection of disjoint paths connecting the taxa
in $S_\gv$ to each other.  Then $q_{\gv}$ is simply the
product of all nontrivial entries of the Fourier transform of the
mutation matrices for edges in $E_\gv$
\cite{sempleSteelPhylogenetics03}. For example, the path collection
corresponding to $\gv = 1010$ is the single path connecting
taxa $1$ and $3$, going through the internal edge. Thus $q_{1010} = \theta
\cdot \theta^{-1} \cdot \theta = \theta$.  All of the other similar
calculations are reported in Table~\ref{tab:ex}. 
% add something about the simple version of FT?
An application
of the inverse Fourier transform gives the $\pv$. Note that because our root
distribution is taken to be uniform, the Fourier transform of the root
distribution is nonzero only at the identity.  Thus the only nonzero
$q_\gv$ are those for which the $\ZZ_2$ sum of the components of $\gv$
equals zero. 

\begin{table}
  \centering
  \begin{tabular}{c|c|c}
    $\gv$ & $q_{\gv}$ & $8 \cdot p_{\gv}$ \\
    \hline
    0000 & $1$ & $1 + 4 \theta + 2 \theta^2 + \theta^4$ \\
    1001 & $\theta$ & $1 - \theta^4$ \\
    0101 & $\theta$ & $1 - \theta^4$ \\
    1100 & $\theta^2$ & $1 - 4 \theta + 2 \theta^2 + \theta^4$ \\
    0011 & $\theta^2$ & $1 - \theta^4$ \\
    1010 & $\theta$ & $(1 - \theta^2)^2$ \\
    0110 & $\theta$ & $(1 - \theta^2)^2$ \\
    1111 & $\theta^4$ & $1 - \theta^4$ \\
  \end{tabular}
  \caption{Site pattern frequencies and their Fourier transforms for
  the example mentioned in the text.}
  \label{tab:ex}
\end{table}

It is clear that in Table~\ref{tab:ex} all $p_{\gv}$ are positive for
$0 \leq \theta \leq 1$ with the possible exception of $p_{1100}$. 
One can ensure positivity of $p_{1100}$ by choosing 
$0 < \theta < 0.2955$, corresponding to a branch length $\gamma >
0.60938$. We fix such a choice of $\theta$, which ensures that $\pv$
sits in the probability simplex. (Note that a less stringent
constraint on the branch lengths could be achieved by taking the
absolute value of the internal branch length to be smaller than the
pendant branch lengths.) Because our $\qv$ comes from Hadamard
conjugation, it satisfies the two phylogenetic invariants in this
setting: $q_{1001} \cdot q_{0110} = q_{1010} \cdot q_{0101}$ and
$q_{0000} \cdot q_{1111} = q_{1100} \cdot q_{0011}$.  Furthermore, the
diagonal entries of the Fourier transform of the mutation matrices
(i.e. $1$, $\theta$ and $\theta^{-1}$) are positive for any real
$\gamma < 0$, and thus the mutation parameters satisfy the two-state
analog of the `biologically meaningful'' criterion of
\cite{casanellasFSK3P07}.  However, this $\qv$ came from a
phylogenetic tree with a negative internal edge. Thus the example begs the question of what
conditions should be put on site-pattern frequency vectors or their
Fourier transforms so that one can be assured that the corresponding
trees are well-formed.

This paper describes the set of ``edge-parameter inequalities'' and shows that
they are the exact conditions needed, namely
that any solution of the phylogenetic
invariants for a given tree which satisfies these inequalities is
guaranteed to come from a tree with non-negative edge parameters. 
For example, an edge-parameter inequality for the
internal edge of the quartet tree is  
\begin{equation}
  q_{0000} \, q_{1111} \, q_{1100} \, q_{0011} \, \geq \,
  q_{1010} \, q_{0101} \, q_{1001} \, q_{0110}
  \label{eq:sampleEPI}
\end{equation}
which is equivalent to the inequality $1 \geq \theta^4$ or
$\gamma \ge 0$. Thus (\ref{eq:sampleEPI}) specifically rules out the
pathological example above.

We will describe two distinct versions of the edge-parameter inequalities. The
first version is derived by considering paths in
the tree and thus will be called the ``path''
edge-parameter inequalities. This version is relatively simple to
write down, involving two monomials
of degree at most four for the two-state models and two monomials of
degree at most six for
the four-state models. We note that as this set of inequalities
is derived on trees, they are only meaningful for $\qv$ which satisfy
a complete set of phylogenetic invariants for a tree.

Next we present the second version of the inequalities; these
inequalities derive
directly from the Sz{\'e}kely-Steel-Erd{\H{o}}s
Fourier conjugation equation \cite{szekelyEAFourierCalc93}. Because
they are given directly by Fourier conjugation, we call these
inequalities the ``canonical'' edge-parameter inequalities. 
These inequalities for group $G$-based models on trees of $m$ taxa
carve out a subset of $q$-space which we denote
$\pep$. The set of $\qv$'s corresponding to a given
$m$-taxon tree is
the set of solutions to that tree's phylogenetic invariants
intersected with $\pep$.

We then investigate some properties of $\pep$. 
The set $\pep$ is the subset of $q$-space which corresponds
precisely to the $\qv$ of splits networks with non-negative split
parameters using an extension of the model of
\cite{bryantExtending05}; thus it is contractible. It is not convex. Furthermore, the
$\qv$ corresponding to phylogenetic trees sit on the boundary of
$\pep$, thus the complete space of phylogenetic ``oranges''
\cite{kimOranges00, moultonSteelOranges04} 
for group-based models lives on this boundary.

Before getting into details, we would like to note that the idea of
constraint inequalities goes back to the remarkable paper of Cavender
and Felsenstein \cite{cavenderFelsensteinSimpleInvars87}.  Indeed,
they anticipate such inequalities, the (phylogenetic) Fourier
transform, and problems with phylogenetic mixtures. Our paper can be
seen as a completion of their investigation of phylogenetic
inequalities for the group-based models.

\section{Technical introduction}

In this section we fix notation and state two versions of Fourier
conjugation. The application of discrete Fourier transform ideas to
phylogenetics was pioneered in
\cite{hendyPennyFramework89,hendyRelationship89} for the CFN model,
then generalized to group-based models in
\cite{szekelyEAFourierCalc93} and 
\cite{evansSpeedInvariants93}.  Our
notation combines that of \cite{sturmfelsSullivantToric05} and
\cite{bryantExtending05}.  We note that because Fourier
conjugation is our primary tool, we will only be considering
group-based mutation models (defined below), in particular $G = \ZZ_2$ or
$\ZZ_2 \times \ZZ_2$.

As stated in the introduction, the simple observation of this paper is
that mutation transition matrices come from continuous-time Markov processes.
Thus the mutation matrices $P^{(e)}$ must satisfy (\ref{eq:PisExpQ})
for each edge 
$e$, with $t_e$ and the off-diagonal elements of $Q^{(e)}$ being
non-negative.
We allow the rate
matrices $Q^{(e)}$ to vary from edge to edge; thus we can (and do)
incorporate $t_e$ into $Q^{(e)}$ and so assume $t_e = 1$ for any $e$.
We call the resulting entries of the mutation rate
matrices $Q^{(e)}$ for an edge ``edge parameters.''
We note that in phylogenetic practice one often assumes a fixed rate
matrix $Q$ for the whole tree and the only parameters of a given edge
are the branch lengths $t_e$; here we make no such restriction.

Fourier conjugation applies to the ``group-based models.'' Each
state in such a model is uniquely labeled with an element of an
Abelian group.
We will write our group $G$ additively, with $0$ denoting the
identity element. 
The essential point in the definition of a group-based model is that
such that the rate of transition from state $g$ to $h$ is
only a function of the difference of $g$ and $h$ in $G$. 
Fixing an edge $e$, we write
\[
Q^{(e)}_{g,h} = \ps{e}{h - g}
\]
where $Q^{(e)}$ denotes the mutation rate matrix along an edge $e$ and
$\psi^{(e)}$ is an arbitrary $|G|$-vector with components summing to zero 
such that $\ps{e}{g} \geq 0$ for $g \neq 0$.
The group-based models considered in the literature are also time reversible, i.e. one
requires that $Q^{(e)}_{g,h} = Q^{(e)}_{h,g}$, which is equivalent to
$\ps{e}{g} = \ps{e}{-g}$.
% what about time reversibility and the root dist'n?
Because exponentiation preserves symmetries of the matrices, we will
also have 
\[
P^{(e)}_{g,h} = \f{e}{h - g}
\]
for some probability $|G|$-vector $f^{(e)}$. Time reversibility similarly implies
$\f{e}{g} = \f{e}{-g}$.

The discrete Fourier transform is constructed via
the ``dual group'' of an Abelian group. The elements of $\dual{G}$, the dual
group to $G$, are the homomorphisms of $G$ to the multiplicative group
of complex numbers of magnitude one. The groups $G$ and $\dual{G}$ are
isomorphic; such an isomorphism is canonical after choosing an
identification of $G$ with a direct product of finite cyclic groups. We
make such a choice, and because of the resulting isomorphism we will
use the same letters $g, h, \dots$ to denote elements of $G$ and
$\dual{G}$.
However, we will follow \cite{bryantExtending05} in using ``hat'' for the
application of an element of the dual group, such that
$\dual{g} (h)$ is the application of $g \in \dual{G}$ to $h \in G$.
(This conflicts with traditional notation for Fourier transform; we
will use ``check'' for this purpose as defined below.)
We also note that because $G$ is isomorphic to a direct
product of cyclic groups we have $\dual{g}(h) = \dual{h}(g)$.

The Fourier transform of a function $a: G \rightarrow \CC$ is
\[
\trans{a}(g) := \sum_{h \in G} \dual{g} (h) a(h).
\]
By the definitions $\ft{e}{0} = 1$ for any $e$. Note
\begin{equation}
  \begin{split}
  \ft{e}{-g} & = \sum_{h \in G} \wdual{-g} (h) \f{e}{h} 
  = \sum_{h \in G} \dual{g} (-h) \f{e}{h} \\
  & = \sum_{h \in G} \dual{g} (h) \f{e}{-h}
  = \sum_{h \in G} \dual{g} (h) \f{e}{h} = \ft{e}{g}
  \end{split}
  \label{eq:ftReal}
\end{equation}
where the fourth equality is by time reversibility.
By the definition of the Fourier transform, $\trans{a}(-g) =
\overline{\trans{a}(g)}$ for any real-valued function $a$. Thus the fact that $\ft{e}{g} = \ft{e}{-g}$
is equivalent to the fact that $\ft{e}{g}$ is real.

The formulas for the phylogenetic Fourier transform are simplified
by re-rooting the tree at a leaf, which
eliminates the need for a special root distribution
\cite{sturmfelsSullivantToric05,szekelyEAFourierCalc93}.  Specifically, we
extend an edge from the root terminating in a new leaf; the previous root distribution is then replaced by a
transition matrix along the new edge. Thus, without loss of
generality, we assume our given tree $\tree$ on $m$ leaves is rooted at a leaf and that the
root distribution puts all mass at the identity.

Phylogenetic Fourier conjugation is an invertible transformation between the
collection of edge parameters $\psi^{(e)}(g)$ and the corresponding
site-pattern frequency vector  for a given tree.  This site-pattern
frequency vector is the joint distribution of states at the leaves
defined as follows. Start at the root, and move towards the leaves,
changing state along an edge $e$ according to $P^{(e)}$. The
induced joint distribution on the leaves will be denoted $\pv$ where
the component $p_{\gv}$ of $\pv$ is the probability of seeing $\gv \in
G^m$ by the above process. 

The Fourier transform of the $\pv$ vector using the group $G^m$ will
be denoted $\qv$. The matrix representation of the Fourier transform
for the group $G$ will be denoted $K$, i.e.  $K_{g,h} := \dual{g}(h)$ for any
$g,h \in G$. The
analogous matrix for $G^m$ will be denoted $H$. Note that $H$ is
the $m$-fold Kronecker product of $K$. In this notation, $\qv = H
\pv$. We note that when $K$ (and thus $H$) is a matrix with entries
$\pm 1$, the Fourier transform is often called the Hadamard transform.

Following \cite{sturmfelsSullivantToric05}, use
$\Lambda(e)$ to denote the set of leaves $i$ such that the path from
$i$ to the root goes through $e$; $\Lambda(e)$ can be thought of the
set of leaves ``below'' $e$. We also define
\[
\upstar g_e = \sum_{i \in \Lambda(e)} g_i.
\]
The vector $\upstar \gv$ is a natural lift of a $\gv \in G^m$ to an
assignment of $G$ to the edges of the tree.
We will be using two versions of Fourier conjugation. 
In this notation, version one can be written
\begin{thm}[Hendy, 1989 \cite{hendyRelationship89}; Evans and Speed, 1993 \cite{evansSpeedInvariants93}]
\begin{equation}
  q_{\gv} = \prod_{e \in E} \ft{e}{\upstar g_e}.
  \label{eq:hadProd}
\end{equation}
  \label{thm:hadProd}
\end{thm}

The second version of the edge-parameter inequalities will use a
different version of the Fourier conjugation.  In order to express
this second version, we state the following lemma:

\begin{lem}
  \label{lem:whatIsEdgeFT}
  \[
  \trans{f}(h) = \exp \left( \trans{\psi}(h) \right).
  \]
\end{lem}
\begin{proof}
  We begin as for Lemma 17.2 of \cite{bryantExtending05} (though for
  right rather than left eigenvalues),
  \begin{equation}
    \begin{split}
      (QK)_{g,h} & = \sum_{x \in G} \psi(x-g) \dual{x} (h)
    = \sum_{y \in G} \psi(y) \wdual{y + g} (h) \\
    & = \dual{g}(h) \sum_{y \in G}
    \psi(y) \dual{y} (h) = K_{g,h} \,\trans{\psi} (h).
  \end{split}
  \label{eq:whatIsEdgeFT}
\end{equation}
  Thus the $h$th column of $K$ is a right eigenvector of $Q$ with eigenvalue
  $\trans{\psi}(h)$. The same argument with $f$ in place of $\psi$ shows
  that the $h$th column of $K$ is a right eigenvector of $P$ with
  eigenvalue $\trans{f}(h)$. However, $P = \exp(Q)$ so the eigenvalues of
  $P$ are the exponentials of the corresponding eigenvalues of $Q$.
\end{proof}

As noted in the discussion after (\ref{eq:ftReal}),
$\trans{\psi}(g)$ is real
for any $g$. Thus Lemma~\ref{lem:whatIsEdgeFT} implies 
\begin{obs}
  \label{obs:FTpos}
  Any edge with real edge parameters will have real and
  non-negative Fourier
  transform $\trans{f}^{(e)}$.
\end{obs}
Thus any tree with non-negative edge parameters has ``biologically
meaningful'' parameters in the language of
\cite{casanellasFSK3P07}, though the converse does not hold. We also note
that by (\ref{eq:hadProd}) the $q_{\gv}$ are
real; thus the logarithm in (\ref{eq:hadCanon}) retains its usual
meaning as a mapping between real numbers.

We will now present a second version of Fourier conjugation. By 
Lemma~\ref{lem:whatIsEdgeFT} and the definition of Fourier transform,
\begin{equation}
  \psi(h) = \left[ K^{-1} \log K f \right]_h
  \label{eq:psiEig}
\end{equation}
where the subscript $h$ denotes the $h$ component of the vector.
The following theorem is
Theorem~6 of \cite{szekelyEAFourierCalc93} in the presence of
(\ref{eq:psiEig}).
\begin{thm}[Sz{\'e}kely, Steel, and Erd{\H{o}}s, 1993]
Let $\rhov(e,h)$ be the element of $G^m$ which assigns $h$ to all
leaves in $\Lambda(e)$ and 0 to all others. Then
\begin{equation}
  \psi^{(e)}(h) = 
  \left[ H^{-1} \log \qv \right]_{\rhov(e,h)}.
    \label{eq:hadCanon}
  \end{equation}
    \label{thm:hadCanon}
\end{thm}
Note that the $\log$ in equation (\ref{eq:hadCanon}) is entry-wise.

\section{Fourier transform inequalities: path version}

In this section we show first that one can very easily extract specific
$\ft{e}{g}$ terms by taking ratios of certain $q_{\gv}$ terms. Then
basic inequalities for the $\ft{e}{g}$ terms will lead to inequalities
in the $q_{\gv}$.
Let
$\path(i,j)$ be the set of edges on the path between nodes $i$ and $j$ in
the tree ($i$ and $j$ may or may not be leaves). Now define
\[
F(i,j;g) = \prod_{e \in \path(i,j)} \ft{e}{g}.
\]
We record the following facts for future use:
\begin{lem}
  \begin{list}{}{}
  \item
  \item[(i)] Let $\nu$ be a node on the path from $i$ to $j$ in a
    tree. Then
    \[ 
    F(i,\nu;g) \cdot F(\nu,j;g) = F(i,j;g).
    \]
  \item[(ii)] $F(i,j;g) = F(j,i;g)$.
  \item[(iii)] $F(i,j;g) = F(i,j;-g)$.
  \end{list}
  \label{lem:F}
\end{lem}
\begin{proof}
  Parts (i) and (ii) are clear from the definition. Equation (\ref{eq:ftReal})
  implies (iii).
\end{proof}

The following fact is a simple application of the above lemma and
Theorem~\ref{thm:hadProd}. 
\begin{lem}
  Let $i$ and $j$ be leaves and let $\gv$ have $g_i = h$,
  $g_j = -h$ and all other components zero. Then
  $q_\gv = F(i,j;h)$.
  \label{lem:path}
\end{lem}

The first identity is for pendant edges. Denote the set of leaves by
$\leaves$.
\begin{prop}
  Given some pendant edge $e$, let $i$ denote the leaf on $e$
  and let $\nu$ be the
  internal node on $e$. Pick $j$ and $k$ any leaves distinct from $i$ such that the path
  $\path(j,k)$ contains $\nu$. Let $w(g_i, g_j, g_k) \in G^\leaves$ assign state
  $g_x$ to leaf $x$ for $x \in \{i,j,k\}$ and the identity to all other leaves. Then
  \begin{equation}
    \left[ \ft{e}{h} \right]^2 = \frac{q_{w(h,-h,0)} \cdot
    q_{w(-h,0,h)}}{q_{w(0,-h,h)}}.
    \label{eq:ftiIntuitivePendant}
  \end{equation}
  \label{prop:ftiIntuitivePendant}
\end{prop}
\begin{proof}
  Lemmas~\ref{lem:F} and \ref{lem:path} show
  \begin{align*}
    q_{w(h,-h,0)} &= \ft{e}{h} \cdot F(\nu,j;h)\\
    q_{w(-h,0,h)} &= \ft{e}{h} \cdot F(\nu,k;h)\\
    q_{w(0,-h,h)} &= F(\nu,j;h) \cdot F(\nu,k;h).
  \end{align*}
\end{proof}

A similar proof implies the next identity, which is for internal
edges.
\begin{prop}
  Pick some internal edge $e$; say the two nodes on either
  side of $e$ are $\nu$ and $\nu'$. Choose $i,j$ (resp. $i',j'$) such
  that $\path(i,j)$ (resp. $\path(i',j')$) contains $\nu$ but not
  $\nu'$ (resp.
  $\nu'$ but not $\nu$). Let $z(g_i, g_j, g_{i'}, g_{j'}) \in G^\leaves$ assign state
  $g_x$ to leaf $x$ for $x \in \{i,j,i',j'\}$ and the identity to all other leaves. Then
  \begin{equation}
    \left[ \ft{e}{h} \right]^2 = 
    \frac{q_{z(h,0,-h,0)} \cdot q_{z(0,-h,0,h)}}
    {q_{z(h,-h,0,0)} \cdot q_{z(0,0,-h,h)}}.
    \label{eq:ftiIntuitiveInternal}
  \end{equation}
  \label{prop:ftiIntuitiveInternal}
  %\qed
\end{prop}

Now, constraints on the $\ft{e}{h}$ will imply 
inequalities in the $q_{\gv}$. Such nontrivial constraints exist; we
review these constraints now for the
usual group based models. 
First we investigate the two-state symmetric (CFN) model, which was described
in the introduction. There is only one non-trivial
component $\ft{e}{1}$ of the Fourier transform along an edge, which is $\exp(-2
\gamma (e))$, where $\gamma(e)$ is the ``branch length'' of that edge.
Now $0 \leq \gamma(e)$ is equivalent to 
\begin{equation}
  \ft{e}{1} \leq 1.
  \label{eq:ftBoundsCFN}
\end{equation}
Inserting the values
for $\ft{e}{1}$ from Propositions~\ref{prop:ftiIntuitivePendant} and
\ref{prop:ftiIntuitiveInternal} into this equation give the
edge-parameter inequalities for each edge. In summary,

\begin{prop}
  \label{prop:JCEdgeIneqs}
  Assume that $\qv$ is the $\ZZ_2$-Fourier transform of a site-pattern
  frequency vector under the CFN model. If \,$\qv$ satisfies a complete
  set of phylogenetic invariants for a tree $\tree$ and a set of inequalities gained by
  substituting an instance of (\ref{eq:ftiIntuitivePendant}) or
  (\ref{eq:ftiIntuitiveInternal}) into the square of (\ref{eq:ftBoundsCFN}) for each
  edge $e$ of \,$\tree$, then $\qv$
  is the expected site-pattern frequency vector of $\tree$ for some
  assignment of non-negative branch lengths to $\tree$.
  Conversely, any tree with non-negative branch lengths will satisfy
  such a set of inequalities.
  %\qed
\end{prop}

As a quick application, we demonstrate how these inequalities exclude the
pathological example described in the introduction. For the internal
edge of this quartet tree under the CFN model, we should have 
\[
\frac{q_{1010} q_{0101}}{q_{1100} q_{0011}} =
\left[ \ft{e}{1} \right]^2 \leq 1.
\]
However, by substituting in values from Table~\ref{tab:ex} the above ratio is
$\theta^{-2}$, which is greater than one. 

For the four-state models we will only discuss the
Kimura three parameter (K3P) model. It is the most general group-based 
four-state model; results for this model extend to less general models
by choosing transition matrices with extra symmetries. The K3P model
is associated with the group $\ZZ_2 \times \ZZ_2$. Thus $K$ for this
model is the Hadamard matrix of order four, which is the
Kronecker product of two Hadamard matrices of order two. 
We make the identifications
\begin{equation}
  A = (0,0) \quad C = (1,0) \quad G = (0,1) \quad T = (1,1).
  \label{eq:K3Pstates}
\end{equation}
We write the column vector $\psi$ as
\[
[-(\psi(C)+\psi(G)+\psi(T)), \psi(C), \psi(G), \psi(T)]^T
\]

Then by Lemma~\ref{lem:whatIsEdgeFT} we have that $\ft{e}{A} = 1$ and
\begin{equation}
  \begin{split}
  \ft{e}{C} & = \exp(-2(\psi(C)+\psi(T))) \\
  \ft{e}{G} & = \exp(-2(\psi(G)+\psi(T))) \\
  \ft{e}{T} & = \exp(-2(\psi(C)+\psi(G))).
  \end{split}
  \label{eq:ftBoundsK3P}
\end{equation}
The following equations are equivalent to requiring 
$\psi(C)$, $\psi(G)$, and $\psi(T)$ to be non-negative via (\ref{eq:ftBoundsK3P}):
\begin{align}
  \label{eq:K3PaIneq}
  \ft{e}{C} \ft{e}{T} & \leq \ft{e}{G} \\
  \label{eq:K3PbIneq}
  \ft{e}{G} \ft{e}{T} & \leq \ft{e}{C} \\
  \label{eq:K3PcIneq}
  \ft{e}{C} \ft{e}{G} & \leq \ft{e}{T}.
\end{align}
In summary, 
\begin{prop}
  \label{prop:K3PedgeIneqs}
  Assume that $\qv$ is the $\ZZ_2 \times \ZZ_2$ Fourier transform of a site-pattern
  frequency vector under the K3P model. If $\qv$ satisfies a complete
  set of phylogenetic invariants for a tree $\tree$ and a set of inequalities gained by
  substituting an instance of (\ref{eq:ftiIntuitivePendant}) or
  (\ref{eq:ftiIntuitiveInternal}) into the square of (\ref{eq:K3PaIneq}),
  (\ref{eq:K3PbIneq}), and (\ref{eq:K3PcIneq}) for each edge $e$ of
  $\tree$, then $\qv$
  is the expected site-pattern frequency vector of $\tree$ for some
  assignment of non-negative edge parameters to $\tree$.
  Conversely, any tree with non-negative edge parameters will satisfy
  such a set of inequalities.
  \qed
\end{prop}

For example, say we substitute (\ref{eq:ftiIntuitiveInternal}) into
the square of (\ref{eq:K3PaIneq}). This gives
\[
\frac{q_{z(C,0,C,0)} \cdot q_{z(0,C,0,C)}}{q_{z(C,C,0,0)} \cdot q_{z(0,0,C,C)}}
\cdot
\frac{q_{z(T,0,T,0)} \cdot q_{z(0,T,0,T)}}{q_{z(T,T,0,0)} \cdot q_{z(0,0,T,T)}}
\leq
\frac{q_{z(G,0,G,0)} \cdot q_{z(0,G,0,G)}}{q_{z(G,G,0,0)} \cdot
q_{z(0,0,G,G)}}
\]
which is equivalent to a monomial inequality of degree six.

Before moving on, we highlight that (\ref{eq:ftiIntuitivePendant}) 
is essentially concerned with induced subtrees on only 3 taxa, and
(\ref{eq:ftiIntuitiveInternal}) is concerned with induced subtrees
on only 4 taxa. Inequalities on the collection of these subtrees imply
positivity of edge parameters for the entire tree.

\section{Fourier transform inequalities: canonical version}
\label{sec:canonical}

The previous section described a relatively simple set of inequalities
which can be computed for any edge of a tree. However, some readers
may feel uncomfortable with the fact that these inequalities involve
some arbitrary choice. In this section we give a ``canonical'' version of
the edge parameter inequalities which is a simple consequence of
Theorem~\ref{thm:hadCanon}. This version of the inequalities also gives a
clearer understanding of the underlying geometry.

We now specialize to the case of either the CFN model or the K3P model
(this again includes K3P with extra symmetries, such as JC DNA and
K2P). In these cases, the entries of the Fourier transform matrix
$K$ are $\pm 1$. 
\begin{prop}
  \label{prop:ftiCanonical}
  Let $G = \ZZ_2$ or $\ZZ_2 \times \ZZ_2$
  and $\rhov(e,h)$ be the element of $G^m$ which assigns $h$ to all
  leaves in $\Lambda(e)$ and 0 to all others. Then for any $\qv$
  generated on a tree with non-negative edge parameters,
  \begin{equation}
    \prod_{\gv: \wdual{\rhov (e,h)} (\gv) = 1} q_\gv \ \geq \ 
    \prod_{\gv: \wdual{\rhov (e,h)} (\gv) = -1} q_\gv
    \label{eq:ftiCanonical}
  \end{equation}
  Conversely, any tree (with edge parameters) whose $\qv$ satisfies
  (\ref{eq:ftiCanonical}) for any $e$ and $h$
  has non-negative edge parameters.
\end{prop}
\begin{proof}
  Recall that $H^{-1} = |G|^m \, H$. Thus (\ref{eq:hadCanon}) is
  \begin{equation}
    |G|^{-m} \, \psi^{(e)}(h) = \left[ H \log \qv \right]_{\rhov(e,h)},
    \label{eq:hadCanonV2}
  \end{equation}
  the left hand side of which is non-negative by our main assumption.
  Exponentiate (\ref{eq:hadCanonV2}); the left hand
  side will be not less than one, and the right hand side becomes a ratio
  with those $q_\gv$ such that $\wdual{\rhov(e,h)} (\gv) = 1$ on top
  and those $q_\gv$ such that $\wdual{\rhov(e,h)} (\gv) = -1$ on
  the bottom. Then multiply to clear denominators.
\end{proof}

Although we have specialized to groups where $K$ has real entries, we
note here that equivalent (though more complex) such inequalities
exist in all cases. First, we
claim that $q_{\hv} = q_{-\hv}$ for any $\hv$. Indeed, assuming time reversibility we
have $f^{(e)}(g) = f^{(e)}(-g)$, thus $q_{\hv} = q_{-\hv}$ by
(\ref{eq:hadProd}). It follows
that the coefficients of the $q_{\hv}$ in $H^{-1} \qv$ are real.
Therefore the same exponentiation process in
Proposition~\ref{prop:ftiCanonical} works, although the $q_{\hv}$ may
now have exponents different than $\pm 1$.

The ``path'' inequalities of Propositions~\ref{prop:ftiIntuitivePendant} and
\ref{prop:ftiIntuitiveInternal}, and the ``canonical''
inequalities of Proposition~\ref{prop:ftiCanonical}, are equivalent.
Indeed, they each express the equation $\psi^{(e)}(h) \geq 0$ for
various $e$ and $h$. However, the expressions are different, but by
the definition of invariants one can go from one to the other formulation via a complete
set of phylogenetic invariants \cite{sturmfelsSullivantToric05}. 

The previous paragraph establishes equivalence between the two
formulations in principle; we present an example here to show how the
transformation works. Assume a quartet tree of topology $12|34$; use
notation as in the introduction. First we investigate the pendant edge
leading to taxon $1$. By (\ref{eq:ftiCanonical}), that edge having
non-negative edge length is equivalent to
\begin{equation}
  q_{0000} \, q_{0110} \, q_{0011} \, q_{0101} \geq
  q_{1100} \, q_{1010} \, q_{1001} \, q_{1111}. 
  \label{eq:pendExampleCanon}
\end{equation}
A couple of algebraic steps using the phylogenetic invariant $q_{1100}
q_{0011} = q_{1111}$ and the fact that $q_{0000} = 1$ 
shows that (\ref{eq:pendExampleCanon}) is equivalent to
\[
1 \ge
\left( \frac{q_{1100} \, q_{1001}}{q_{0101}} \right)
\left( \frac{q_{1100} \, q_{1010}}{q_{0110}} \right),
\]
which is the product of the two ``path'' pendant edge length
inequalities. Similarly, the internal edge being non-negative is
equivalent to
\[
1 \ge
\frac{q_{1010} \, q_{0101} \, q_{1001} \, q_{0110}}
     {q_{0000} \, q_{1111} \, q_{1100} \, q_{0011}} = 
\left( \frac{q_{1010} \, q_{0101}}{q_{1100} \, q_{0011}} \right)
\left( \frac{q_{1001} \, q_{0110}}{q_{1100} \, q_{0011}} \right)
\]
where the right hand side of the equality is the product of the two
``path'' internal edge length inequalities.

The canonical construction generalizes the inequalities to the
more general setting of group-based mutation models on split networks as
formulated by David Bryant \cite{bryantExtending05}. Assume the set of
splits is labeled $\Sigma$.
In his elegant formulation, one assigns mutation
probabilities to each possible split, i.e. a probability distribution
on the group $G$ for each split. Assuming independence of these
distributions, one gets a probability distribution on $G^{\Sigma}$ by
multiplication. 
From there the probability of a single
site-pattern $\hv$ (i.e. the assignment of a group element to each taxon) is
the sum of the probabilities of all elements of $G^{\Sigma}$ which
give $\hv$ on the leaves.

Fourier conjugation also works in this setting. Although
Bryant's paper \cite{bryantExtending05} only develops the conjugation in the case of
models with a fixed rate matrix and ``branch length'' varying among
splits, there is also an invertible transformation for the setting where one
allows the whole rate matrix to vary. We will apply this extended
version and call the set
of $\psi^{(e)}$ for splits $e$ ``split parameters'' analogous to the
edge parameters we have been describing so far.
Although we do not go into details here, the proof of the Fourier
conjugation formula in the extended case is similar to that in
\cite{bryantExtending05}.  One can then obtain an equation for the
Fourier conjugation written exactly as in (\ref{eq:hadCanon}) but with
a generalized definition of the terms:  ``root'' the
splits network at the taxon $n$, and so redefine $\Lambda(e)$ to be
all of the taxa on the opposite ``side'' of the split from $n$. For
example, $\Lambda(12|34)$ is the set $\{1,2\}$ as in this case $n=4$.

\begin{defn}
  Let $\pep$ be the points of $q$-space which satisfy inequalities
  (\ref{eq:ftiCanonical}) for each split $e$ and each $h \in G$.
\end{defn}

\begin{obs}
  \ \newline
  \vspace{-0.4cm}
  \begin{enumerate}[(i)]
    \item $\pep$ is the image of the non-negative split parameter splits
  networks under Hadamard conjugation.
    \item $\pep$ is contractible.
    \item The points of $q$-space corresponding to trees of topology $\tree$ with
      non-negative edge
      parameters are the zero set of the phylogenetic invariants for
      $\tree$ intersected with $\pep$. These points sit on the boundary of
      $\pep$ for $m > 3$.
  \end{enumerate}
\end{obs}

\begin{proof}
  We note that $\pep$ is the (injective) image of the
  set of non-negative split parameter vectors in 
  $(\RR_{\ge 0})^{2^{m-1} \cdot (|G|-1)}$.  
  For (i), the inequalities (\ref{eq:ftiCanonical}) precisely specify
  positivity of split parameters.
  For (ii) the required homotopy simply uniformly shrinks every split parameter
  to zero.
  The first sentence of (iii) is
  equivalent to Proposition~\ref{prop:ftiCanonical}. For the second
  sentence, the boundary of $\pep$ consists of the image of splits
  networks with at least one zero split parameter. Phylogenetic trees
  are simply split networks such that only a compatible set of split
  parameters are nonzero.
\end{proof}

This series of observations suggests that rather than
phylogenetic ``orange'' \cite{kimOranges00} with one orange slice
for each tree topology, one might think of a phylogenetic ``soccer
ball'' with one panel of the soccer ball for each tree topology. Indeed,
the set of Fourier transformed points corresponding to any tree live
on the boundary of a higher dimensional contractible object. However,
it should be noted that not every point of the boundary of $\pep$ 
corresponds to a phylogenetic tree, and in fact the panels are of
strictly lower dimension than the boundary of the soccer ball.

Furthermore, we now show that the soccer ball $\pep$ is not convex.
Recall that $\ft{e}{g}$ is real by the discussion after
(\ref{eq:ftReal}). Then:
\begin{lem}
  \label{lem:fHatLessOne}
  The components of the Fourier transformed mutation probability
  vector $\ft{e}{g}$ are
  less than or equal to one for any edge $e$ with non-negative
  edge parameters.
\end{lem}
\begin{proof}
  By Lemma~\ref{lem:whatIsEdgeFT}, it suffices to show that
  $\trans{\psi}^{(e)}(g)$ is non-positive. By the
  definition of $\psi$,
  \[
  \psi(0) = - \sum_{g \neq 0} \psi(g)
  \]
  which implies that $\trans{\psi}^{(e)}(g)$ is non-positive by the definition
  of the discrete Fourier transform.
\end{proof}

\begin{prop}
  \label{prop:pepNonConvex}
  $\pep$ is not convex for $m \geq 3$ and $G = \ZZ_2$ or 
  $\ZZ_2 \times \ZZ_2$.
\end{prop}
\begin{proof}
  We report the argument for the case of $G = \ZZ_2 \times \ZZ_2$
  (i.e. K3P); the case of $G = \ZZ_2$ is analogous but easier.
  We label the sates $A, C, G, T$ as in (\ref{eq:K3Pstates}).
  Pick an arbitrary tree $\tree$ on $m$ taxa; Find a cherry (two-taxon
  rooted subtree) of $\tree$ and label the leaves of $\tree$ with $1,2$.
  Number the edge leading to taxon $1$ with $1$, the edge leading to
  taxon $2$ with $2$, and the edge meeting $1$ and $2$ with $3$. Pick
  arbitrary $0 \leq \theta_1, \theta_2, \theta_3 \leq 1$ such that
  \begin{equation}
    \theta_1 \theta_2 < \theta_3^2 \left( (\theta_1 + \theta_2) / 2
    \right)^2;
    \label{eq:badParams}
  \end{equation}
  this is easily achieved by fixing $\theta_2$ and $\theta_3$ and
  taking $\theta_1$ to be small.

  We will construct two vectors $\qv',\qv'' \in \pep$ such that
  $\qv := (\qv' + \qv'')/2$ is not in $\pep$. The vectors $\qv'$ and
  $\qv''$ will be defined via the Fourier transform by specifying
  their $\ft{e}{g}$. It can be checked that $\qv'$ and $\qv''$ sit in
  $\pep$ using Lemma~\ref{lem:whatIsEdgeFT}, then taking the logarithm and
  the inverse Fourier transform.
  
  Let $V = \{C, T\}$. For $\qv'$ set 
  \[
  \ft{1}{g} = \theta_1 \quad 
  \ft{2}{g} = \theta_2 \quad 
  \ft{3}{g} = \theta_3 
  \]
  for $g \in V$, and $\ft{e}{g} = 1$ otherwise. For $\qv''$ set
 \[
  \ft{1}{g} = \theta_2 \quad 
  \ft{2}{g} = \theta_1 \quad 
  \ft{3}{g} = \theta_3 
  \]
  for $g \in V$, and $\ft{e}{g} = 1$ otherwise. 

  We claim that $\qv$ violates (\ref{eq:ftiCanonical})
  with $e = 3$ and $h = C$, and thus does not sit in $\pep$. To establish this claim, we calculate each
  side of (\ref{eq:ftiCanonical}). First note that $\dual{C} (g) = -1$
  for $g \in V$ and is 1 otherwise. 
  Thus $\wdual{\rhov(3,C)}(\gv) = -1$ exactly when 
  $|\{g_1, g_2\} \cap V|$ is odd, and is $1$ otherwise (here and below
  the notation $g_i$ denotes the $i$th-taxon component of $\gv$). 

  Define $q_{u(x_1, x_2)}$ to be $q_\gv$ for any $\gv$ such that $g_1
  = x_1$ and $g_2 = x_2$. This $q_{u(x_1, x_2)}$ is well defined via
  (\ref{eq:hadProd}) because all $\ft{e}{g} = 1$ except when $e =
  1,2,3$.  Noting that $C+C = 0$, we see that $q_{u(C,C)} = \theta_1
  \theta_2$ by (\ref{eq:hadProd}). Similarly, \[ q_{u(C,A)} =
  q_{u(A,C)} = \theta_3 (\theta_1 + \theta_2 ) / 2.  \]
  
  Because we have arranged that 
  $\trans{f}^{(e)}(A) = \trans{f}^{(e)}(G)=1$ and 
  $\trans{f}^{(e)}(C) = \trans{f}^{(e)}(T)$ for both $\qv'$ 
  and $\qv''$, there are three cases for $q_{u(x_1,
  x_2)}$.  If $x_1$ and $x_2$ are in $V$ then $q_{u(x_1, x_2)} =
  q_{u(C, C)}.$ If $|\{x_1, x_2\} \cap V|$ is one then $q_{u(x_1,
  x_2)} = q_{u(C, A)}.$ If neither $x_1$ nor $x_2$ are in $V$ then
  $q_{u(x_1, x_2)} = 1.$
  
  Thus (\ref{eq:ftiCanonical}) is in this case 
  \[
  \left(q_{u(C,C)}^4\right)^{4^{m-2}} \geq \left( q_{u(C,A)}^8
  \right)^{4^{m-2}}.  
  \] 
  Taking both sides to the power of $4^{1-m}$
  and substituting gives 
  \[ 
  \theta_1 \theta_2 \geq \theta_3^2 \left(
  (\theta_1 + \theta_2) / 2 \right)^2, \] violating
  (\ref{eq:badParams}).  \end{proof}

Proposition~\ref{prop:pepNonConvex} has an interesting phylogenetic
interpretation along the lines of \cite{matsenSteelMimic07}: there are
mixtures of two site pattern frequency vectors corresponding to trees
such that the splits network corresponding to the mixture has 
negative edge parameters. However, the trees used in the proof had
many edge-parameters zero; this is not strictly necessary though it
greatly simplifies the proof.

\section{Consequences and Conclusions}

In summary, we have presented a collection of inequalities in the
Fourier transformed site-pattern frequency space which are equivalent
to the
assumption that group-based mutation rate matrices have non-negative off-diagonal
entries. We are motivated in part by the idea of formulating maximum
likelihood as a constrained optimization problem
\cite{hostenEASolvingLikelihood05,sturmfelsSullivantToric05}. We noted
in the introduction that the previously known constraints are not sufficient to
ensure that the result of the constrained optimization is in fact a
proper tree.  As described in Propositions~\ref{prop:JCEdgeIneqs},
\ref{prop:K3PedgeIneqs}, and \ref{prop:ftiCanonical}, our
inequalities complete the set of constraints: if
a $\qv$ satisfies a complete set of phylogenetic invariants and
the inequalities described here, then it does indeed correspond to a
\emph{bona fide} tree. Thus phylogenetic invariants along with the
edge-parameter inequalities could indeed be safely used to formulate
maximum-likelihood phylogenetic estimation as a constrained
optimization problem.

We also defined $\pep$, which is the set of $\qv$ which come
from splits networks with non-negative edge parameters. We noted
that the tree images for each tree topology sit on the
boundary of $\pep$. Here we showed that $\pep$ is not convex at a
number of points. Note that because $\pep$ is cut out by
monomial inequalities (\ref{eq:ftiCanonical}) one would expect that
$\pep$ would be non-convex at ``most'' points.

As the edge-parameter inequalities are the second component of the
constraints for phylogenetic trees, one might wonder if they could be
used for phylogenetic inference in an manner analogous to phylogenetic
invariants \cite{cavenderFelsensteinSimpleInvars87,erikssonUsingInvariants07}.
In a sense these inequalities appear more natural than phylogenetic
invariants for the purpose of determining the tree corresponding to a
data set: given a real-world data set, one might actually hope that
the
inequalities presented here could be satisfied, whereas phylogenetic
invariants (which are equalities) will essentially never be. Using the
terminology above, one might hope that data would sit in the interior of
$\pep$ even though one would never expect data to sit on its boundary.

This hope is not justified for simulated data on a
tree. Indeed, one can think of the simulated data points as some
distribution centered on the expected distribution. Recall that the
set of trees are simply the set of splits networks with some edge
parameters set to zero.  If the simulation distribution does not have
support on some lower-dimensional surface, then the pre-image of the
distribution will almost certainly have points with negative
coordinates in parameter space. Said another way, it is improbable 
that a sample from a distribution centered on a ``corner'' of the
boundary of $\pep$ would sit in the interior of $\pep$. As an example one might look at
Figure~17.1 of \cite{felsensteinInferringPhylogenies04} where negative
split parameters (besides that for the trivial split) are encountered
in a simulation. Despite these challenges, edge-parameter inequalities
may well prove useful for inference.

We acknowledge that all of the work presented here is for group-based models.
This is a rather strong restriction as all group-based models must
have uniform stationary distribution; real data sets
rarely have this feature. Presumably, there are inequalities
corresponding to those presented here for non-group based models.
However, as no Fourier transform is available for those models the
formulation may be very complex.

\section*{Acknowledgments}
The starting point for this paper came from discussions with a number
of very helpful people, including Elizabeth Allman, David Bryant, Mike
Steel, and Bernd Sturmfels. The author would like to thank Seth
Sullivant for clearing up a technical point from their paper, Steve
Evans and Lior Pachter for comments on an early version of the
manuscript, and Dustin Cartwright for several helpful discussions.
F.A.M. is funded by the Miller Institute for Basic Research
in Science at the University of California, Berkeley.

\bibliographystyle{IEEEtran.bst}
\bibliography{ft_ineqs}

\end{document}